\documentstyle[aps,prl,epsf,multicol]{revtex}
\draft
\begin{document}
\title{The Hoop Conjecture and Cosmic Censorship in the Brane-World}
\author{Ken-ichi Nakao$^{1}$, Kouji Nakamura$^{2}$ and 
  Takashi Mishima$^{3}$
} 
\address{
$^{1}$Department of Physics, Graduate School of Science, 
Osaka City University, Osaka 558-8585, Japan
}
\address{
$^{2}$Division of Theoretical Astrophysics,~National
Astronomical Observatory,~Mitaka, Tokyo 181-8588,~Japan
}
\address{
$^{3}$Laboratories of Physics,~College of Science and Technology,~
Nihon University,\\ Narashinodai,~Funabashi,~Chiba 274-0063,~Japan
}
\date{\today}
\maketitle
\begin{abstract}

The initial data of gravity for a cylindrical matter
distribution confined on the brane is studied in the framework
of the single brane Randall-Sundrum scenario.
We numerically found that the sufficiently thin configuration of matter 
leads to the formation of the marginal surface on the brane 
in the Randall-Sundrum model, 
even if the configuration is infinitely long. 
This means that the hoop conjecture proposed by Thorne 
does not hold in the Randall-Sundrum scenario; 
Even if a mass $M$ does not get compacted into a region whose 
circumference$\,(C)$ in every direction is 
$C 
\lesssim 
4\pi GM $, 
black holes with horizons can form on the brane-world of 
the Randall-Sundrum scenario.

\end{abstract}      

\begin{multicols}{2}

Assuming 4-dimensional general relativity and
physically reasonable conditions on the matter fields,  
Thorne has proven that there is no marginal surface in the
system of a cylindrical distribution of matter fields\cite{Ref:HOOP}. 
In his proof, the marginal surface means a cylindrically
symmetric spacelike 2-surface such that the expansion of the
outgoing null normal to this surface vanishes. 
This result, together with the Newtonian analogy, has led to the
so called hoop conjecture for the necessary and sufficient
condition on black hole formation; 
{\it Black holes with horizons form when and only when a mass
  $M$ gets compacted into a region whose circumference in every
  direction is ${C}\lesssim 4\pi GM$}\cite{Ref:HOOP}.
The converse of the hoop conjecture gives a criterion of
naked singularity formation; 
{\it If a mass $M$ forms a singularity but does not get
  compacted into a region whose circumference in every direction
  is ${C}\lesssim 4\pi GM$, then the singularity will be naked.} 
Then the hoop conjecture is closely related to the cosmic
censorship proposed by Penrose\cite{Ref:penrose69} which states, 
roughly speaking, that the gravitational collapse in a physically
reasonable situation does not lead to naked singularities. 
If the hoop conjecture is really true in our universe, and if a
singularity formed by physically reasonable gravitational
collapse is not confined within a region of 
$C\lesssim 4\pi GM$, it is a counter example for the cosmic
censorship.

No counter example for the hoop conjecture has been presented.
Indeed, some numerical works have been done to confirm the hoop
conjecture.
The numerical simulations by Nakamura et al. strongly suggest
that a highly elongated axisymmetric cold fluid forms a spindle
naked singularity\cite{Ref:NAKED1}. 
Later, Shapiro and Teukolsky also showed that the same is true
for collisionless particle systems\cite{Ref:NAKED2}.
These numerical works suggest that the hoop conjecture will be
true in 4-dimensional general relativity.

However, strictly speaking, we do not know whether general
relativity can describe strong gravity in our real 
universe even for classical situations. 
We have no experimental evidence for it. 
If general relativity is inapplicable to the situation of the
strong gravity, it again becomes a non-trivial issue 
whether a highly elongated spindle gravitational collapse could
form naked singularities or not.

As an alternative theory of gravity, Randall and Sundrum (RS)
recently proposed a scenario of the compactification of a higher
dimension without compact manifold\cite{Ref:RS2}. 
They considered 5-dimensional spacetimes with negative
cosmological constant $\Lambda<0$ including a single 
3-brane with the positive tension $\lambda$.
In their scenario, all the physical fields, except for 
gravity, are assumed to be confined on the RS brane and the gravity
is governed by 5-dimensional Einstein gravity.
Using the fine tuning $\sqrt{-6/\Lambda} = 1/\lambda =: l$,
they showed that even without a gap in the Kaluza-Klein spectrum, 
4-dimensional Newtonian and general relativistic gravity on the
brane is reproduced to more than adequate precision\cite{Ref:RS2,Ref:SSM}. 
The deviation in the gravitational force from the Newtonian one
appears in the short scale less than $l$\cite{Ref:RS2,Ref:GT}. 
Since experimental tests have already proved the $1/r^{2}$
corrections to the Newtonian gravitational potential up to the
sub-millimeter order\cite{Ref:Newton-Correction}, 
the length scale of $l$ must be less than millimeter scale.

In the RS scenario, the cancellation of the long range forces due
to the negative cosmological constant and the brane tension 
reproduces the 4-dimensional gravity on the RS brane. 
However the sufficiently short range gravity(about less than the scale $l$) 
seems to be so insensitive to the cosmological constant 
that the 5-dimensional nature of the gravity may appear and become 
important for the formation of black holes on the RS brane. 
In fact within the limits of linear analysis, it has been proven that 
the gravity in RS model at short distance is 
5-dimensional~(see e.g.\cite{Ref:GKR}). 
If the results of the linear analysis can apply to the problem of 
the formation of black holes on the RS brane, 
we may imagine the possibility of thin spindle-like black holes and 
even infinitely long ones on the RS brane, 
considering the existence of black string solutions 
in 5-dimensional Einstein gravity~\cite{Ref:GL}. 
Then we are lead to the guess that the hoop conjecture on the RS brane 
may not be valid in the scale less than $l$.

Of course this discussion should be justified by the fully non-linear
analyses, since the formation of a black hole is a fully non-linear
phenomenon. 
Moreover in the case of the cylindrical matter distribution on the RS
mode, two different types of singular sources (a singular 3-brane and
a singular matter distribution on it) interact with each other in
complex way. 
Then the general relativistic analysis without any weak field
assumption is necessary to confirm the validity of the hoop
conjecture on the RS brane.

In order to get an insight into the black hole formation 
and the hoop conjecture in the RS scenario, 
we consider a cylindrically symmetric matter distribution on the brane 
and formation of marginal surfaces. 
In this letter, we concentrate a time-symmetric initial data 
which is a 4-dimensional spacelike hypersurface embedded in the 
whole spacetime with vanishing extrinsic curvature\cite{Ref:SS} 
as a first step. The line element of the hypersurface we adopt here is 
\begin{equation}
  d\ell^2 = \phi^2(R,\xi)\,
            \left[ dR^2+R^2d\varphi^2+dz^2+e^{2\xi/l}d\xi^2 \right],
  \label{eq:line-element}
\end{equation}
where the coordinate $\xi$ corresponds to the extra dimension.

The conformal factor $\phi$ which is determined by the initial value
constraints of 5-dimensional Einstein gravity, i.e., the
Hamiltonian constraint and momentum constraints. 
Since we concentrate on the time symmetric initial data, 
the momentum constraints are satisfied trivially. 
On the other hand, the Hamiltonian constraint is given by 
\begin{eqnarray}
  &\biggl\{&\partial_{R}^{2}+{1\over R}\partial_{R}
  +e^{-2\xi/l}\left(\partial_{\xi}^{2}
    -{1\over l}\partial_{\xi}\right)\biggr\}\Omega \nonumber \\
  && \quad\quad -{2\over l^{2}}\Omega\left(\Omega^{2}+3e^{-\xi/l}\Omega
    +3e^{-2\xi/l}
  \right)=0, \label{eq:constraint-1}
\end{eqnarray}
where we have introduced a new variable defined by 
$\Omega:= \phi-e^{-\xi/l}$.

The existence of the brane and the matter fields is taken into
account by the boundary condition at $\xi=0$. 
By the Israel's prescription\cite{Ref:Israel}, we can easily see
that the extrinsic curvature of the brane in the initial
hypersurface has discontinuity which is related to the tension
of the brane and the stress-energy tensor of the matter fields.
Following the RS model, we impose $Z_{2}$-symmetry with respect
to $\xi=0$\cite{Ref:yamaguchi-kitazawa}.
Then the boundary condition on the brane $\xi=0$ is given by
\begin{eqnarray}
\partial_{\xi}\left(e^{2\xi/l}\Omega\right)
+{\Omega^{2}\over l}
+{4\over3}\pi G_{5} 
\rho(R)(\Omega+1)^{2} = 0, \label{eq:B-condition-1}
\end{eqnarray}
where $G_{5}=Gl$ is the 5-dimensional Newton's gravitational
constant\cite{Ref:Emparan}, 
$\rho(R)$ is the energy density of the matter fields.
The 4-dimensional Minkowski spacetime $(\Omega=0)$ is realized 
on the brane when $\rho(R)=0$.

Because of the ambiguity in the original statement by Thorne,
there are many proposals and attempts to prove the hoop
conjecture\cite{Ref:hoop-misc}. 
In this letter, we concentrate on the situation in which the support of
the energy density $\rho(R)$ is in the infinite cylinder with the
coordinate radius $R_{\rm s} $.
Further, as a definition of the mass, we adopt the total
proper mass
\begin{equation}
  M := 2\pi \int_{0}^{R_{\rm s} } dR R \int_{-L}^{L} dz \phi^{3}(R,0) \rho(R)
  \label{eq:mass-def}
\end{equation}
within the cylinder of the coordinate radius $R_{\rm s} $ and a
finite coordinate length $2L$.
As the definition of the circumference, we adopt the proper
length
\begin{equation}
  C := 2 \int_{-L}^{L}dz\phi(R_{\rm s} ,0) 
  + 4 \int_{0}^{R_{\rm s} }dR\phi(R,0)
  \label{eq:circum-def}
\end{equation}
of this cylinder.

We are interested in the formation of a marginal surface in
  a situation of $C > 4\pi G M$, and
can easily check that the inequality $C>4\pi GM$  
holds for arbitrary $L$ if and only if 
the following inequality is satisfied: 
\begin{equation}
1 \geq 
{4\pi^{2}G_{5}\over l\phi(R_{\rm s},0)}
\int_{0}^{R_{\rm s}}dRR\phi^{3}(R,0)\rho(R).
  \label{eq:hoop-criterion}
\end{equation}
It should be noted that this inequality gives an upper 
bound on the line energy density of the cylindrical matter field.

If the existence of future null infinity similar to the AdS 
or Minkowski spacetime and further the global 
hyperbolicity in the causal past of the future null infinity 
are guaranteed, the formation of a marginal surface in the 
initial data might mean the formation of a black hole with horizon. 
Then we may regard that the initial data, in which there is a
marginal surface and the inequality (\ref{eq:hoop-criterion})
holds, as a counter example of the hoop conjecture.

There are two kinds of the marginal surfaces on the initial
data: one is defined by the ``null'' rays confined in the
brane and the other is defined by the null rays which 
propagate in the whole spacetime including the bulk. 
We call these two marginal surfaces {\it Brane-MS} and 
{\it Bulk-MS}, respectively. 
As commented in Ref.\cite{Ref:SS}, these two marginal surfaces
have different physical meanings from each other.
Brane-MS is the marginal surface for all the physical fields
confined on the brane. On the other hand, Bulk-MS is for 
all the physical fields including 
gravitons which propagate in the whole 5-dimensional spacetime. 
Brane-MS is not concerned with the causal structure of the whole
spacetime but Bulk-MS is.

Brane-MS for the cylindrical matter distribution is a
cylindrical 2-surface specified by $R=$constant on the brane
where the expansion of the null rays normal to the surface 
vanishes.
In this case, the null rays are 
defined by the induced metric on the brane. 
On the time symmetric initial hypersurface, the condition of
vanishing expansion of these null rays is equivalent to 
$\partial_{R}\left(R\phi^{2}\right)|_{\xi=0}=0$. 
This is the equation for the coordinate radius of the Brane-MS.

On the other hand, Bulk-MS for the cylindrical matter
distribution is a cylindrical spacelike 3-surface defined in the
whole of a 4-dimensional spacelike hypersurface, on which the 
expansion of the outgoing null
rays normal to the 3-surface vanishes.  
In this case, the null rays defined by the metric on the
whole spacetime.
On the time symmetric initial hypersurface, Bulk-MS is expressed
by a curve in $(R,\xi)$-plane,  which intersects the axes $R=0$
and $\xi=0$.
To find this curve, we introduce spherical polar coordinate
variables $r:= \sqrt{R^{2} + l^{2}(e^{\xi/l}-1)^{2}}$ 
and $\theta:= \tan^{-1}\{R/l(e^{\xi/l}-1)\}$. 
Then a cylindrical spacelike 3-surface will be specified by the
function $r=r(\theta)$.  
The condition of the vanishing expansion leads to an ordinary
differential equation for $r(\theta)$ as
\begin{eqnarray}
{d^{2}r\over d\theta^{2}}&=& 
-{3\over r^{2}}\left({dr\over d\theta}\right)^{3}\partial_{\theta}\ln\phi
+\left({dr\over d\theta}\right)^{2}
\left({2\over r}+3\partial_{r}\ln\phi\right) \nonumber\\  
&-&3{dr\over d\theta}\partial_{\theta}\ln\phi 
+r+3r^{2}\partial_{r}\ln \phi \nonumber \\
&+&r\left\{1+{1\over r^{2}}\left({dr\over d\theta}\right)^{2}\right\}
\left(1-{1\over r}\cot\theta{dr\over d\theta}
\right). \label{eq:MS-equation}
\end{eqnarray}
The boundary conditions both at $\theta=0$ and at $\pi/2$ to 
this equation are given by $dr/d\theta=0$. 
We search for solutions of Eq.(\ref{eq:MS-equation}) 
numerically by the shooting method.

We solve Eq.(\ref{eq:constraint-1}) numerically by the finite  
difference method. The numerically covered region is 
$R_{\rm max}\geq R \geq 0$ and $\xi_{\rm max}\geq \xi \geq 0$. 
Then we need to specify the boundary conditions at 
four kinds of numerical boundaries; $R=0$, $R=R_{\rm max}$, $\xi=0$ 
and $\xi=\xi_{\rm max}$. We impose $\partial_{R}\Omega=0$ at 
$R=0$ and Eq.(\ref{eq:B-condition-1}) 
at $\xi=0$. To fix the boundary conditions at $R=R_{\rm max}$ and 
at $\xi=\xi_{\rm max}$, we assume that the system is isolated, 
i.e., the space approaches to that of the AdS spacetime for 
$r\rightarrow\infty$.
Therefore, the solution should behave as those to linearized
Eq.(\ref{eq:constraint-1}) near the numerical boundaries
$R=R_{\rm max}$ and $\xi=\xi_{\rm max}$.

The linear solution to Eq.(\ref{eq:constraint-1}) is obtained as
follows:
Regarding $|\Omega|= O(G_{5}l\rho) \ll 1$, we derive the
linearized version of Eqs.(\ref{eq:constraint-1}) and
(\ref{eq:B-condition-1}) and solve this linearized equation with
a singular line source
$\rho(R)=\sigma_{\rm L}\delta(R)/2\pi R$, 
where $\sigma_{\rm L}$ is the line energy density of the
line source.
The linear solution $\Omega=\Omega_{\rm L}$ is then given by 
\begin{eqnarray}
& &\Omega_{\rm L}(R,\xi) := 
- {2G_{5}\sigma_{\rm L}\over l} \biggl\{e^{-2\xi/l} 
\ln \left({R\over R_{\rm c}}\right) \nonumber \\
& &\qquad\qquad\qquad\qquad\qquad
-{l\over3} 
\int_{0}^{\infty}dmu_{m}(l)u_{m}(le^{\xi/l})K_{0}(mR)\biggr\},
\label{eq:L-solution}
\end{eqnarray}
where $R_{\rm c}$ is an integration constant which corresponds to 
the freedom to rescale $\xi$, $K_{0}(x)$ is the modified 
Bessel function of the 0th kind, and $u_{m}(\psi)$ is a
combination of the spherical Bessel functions of the first and
second kinds;
\begin{eqnarray}
u_{m}(\psi) &=& \sqrt{\frac{2(ml)^{4}}{\pi((ml)^{2}+1)}}\nonumber\\
&\times&
m\psi(n_{1}(ml)j_{2}(m\psi)-j_{1}(ml)n_{2}(m\psi)).
\end{eqnarray}

Using the linear solution (\ref{eq:L-solution}), we impose the
following boundary conditions at these numerical boundaries: 
\begin{eqnarray}
\partial_{R}\left\{{\Omega(R,\xi)\over \Omega_{\rm L}(R,\xi)}\right\}&=&0;
~~~~~{\rm at}~~R=R_{\rm max}, \label{eq:B-condition-3}\\
\partial_{\xi}\left\{{\Omega(R,\xi)\over \Omega_{\rm L}(R,\xi)}\right\}&=&0,
~~~~~{\rm at}~~\xi=\xi_{\rm max}.\label{eq:B-condition-4}
\end{eqnarray}
These boundary conditions guarantee that the numerical solution 
$\Omega$ behaves as the linear solution 
$\Omega_{\rm L}$ near the numerical
boundaries $R=R_{\rm max}$ and $\xi=\xi_{\rm max}$.

To guarantee that our cylindrical matter 
distribution is an isolated system, further, 
$|\Omega|$ should be much smaller than unity in the vicinity of these 
numerical boundaries. However the boundary conditions 
(\ref{eq:B-condition-3}) and (\ref{eq:B-condition-4}) impose no 
constraint on the amplitude of $\Omega$ itself at these boundaries and hence 
these boundary conditions are not sufficient. This is accomplished 
by choosing the line energy density of the system considered here 
to be sufficiently small.  
  As will be shown below, this is naturally realized in a situation 
  in which the inequality (\ref{eq:hoop-criterion}), or equivalently, 
  $C > 4\pi GM$ holds.

To derive the numerical solutions, we consider the following
energy density $\rho(R)$; for $R<R_{\rm s}$ 
\begin{equation}
\rho(R)(\Omega+1)^{2}|_{\xi=0}={3\sigma\over \pi R_{\rm s}^{2}}
\left\{\left({R\over R_{\rm s}}\right)^{2}-1\right\}^{2},
\label{eq:Non-singular-rho}
\end{equation}
while $\rho(R)=0$ elsewhere, where $\sigma$ is a constant 
which corresponds to the line energy density of this system. 
We choose $l=1$ (this is regarded as a choice of the unit) and then set 
the numerical boundaries to be $R_{\rm max}=\xi_{\rm max}=2$. 
The number of numerical grids for Eq.(\ref{eq:constraint-1}) 
is $300\times 300$, while that for Eq.(\ref{eq:MS-equation}) is $100$. 
Further, we chose $G_{5}\sigma=3\times 10^{-2}n$ and 
$R_{\rm s}=G_{5}\sigma/6=5\times10^{-3}n$, where $n=1,~2,~3$. 
This choice of $\sigma$ 
guarantees $|\Omega|\ll 1$ at the numerical boundaries $R=R_{\rm max}$ 
and $\xi=\xi_{\rm max}$.

We have checked our numerical codes for the initial data and for
marginal surfaces by comparing the $l=\infty$ numerical solution of the
energy density (\ref{eq:Non-singular-rho}) with the $l=\infty$
analytic solution
\begin{equation}
\Omega(R,\xi)={2G_{5}\sigma\over 3 r}. \label{eq:V-solution}
\end{equation}
The solution (\ref{eq:V-solution}) is obtained by solving the
Hamiltonian constraint (\ref{eq:constraint-1}) with the boundary
condition (\ref{eq:B-condition-1}) with a singular line source
$\rho(R)(\Omega+1)^{2}=\sigma\delta(R)/(2\pi R)$ under the
situation $l\rightarrow\infty$.

\begin{figure}
  \begin{center}
    \leavevmode
    \epsfxsize=0.45\textwidth
    \epsfbox{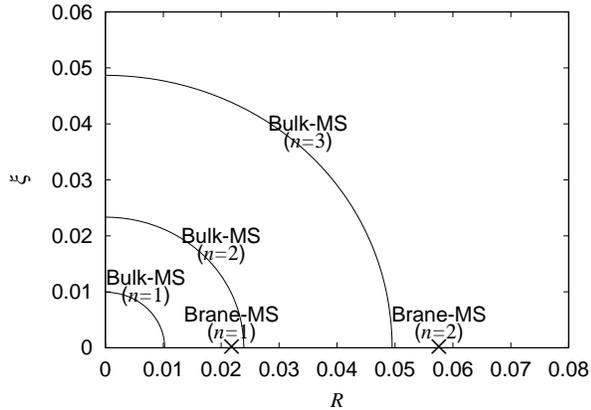}    
  \caption{ 
    We set $G_{5}\sigma=3\times10^{-2}n$ and $l=1$. Then  
    Bulk-MS of $n=1,~2,~3$ and the location 
    of Brane-MS of $n=1,~2$ are depicted. 
    Brane-MS of $n=3$ is located outside this figure ($R\sim0.132$).
    The matter is located within $0\leq R <G_{5}\sigma/6=5\times10^{-3}n$
    at $\xi=0$. From this figure, we can see that $n=1$ case 
    almost agree with the $l=\infty$ analytic solution; 
    Bulk- and Brane-MS of $l=\infty$ are given by $r=G_{5}\sigma/3$ 
    and $R=2G_{5}\sigma/3$, respectively. }
 \label{fig:M-surface}
  \end{center}
 \end{figure}

We found both Brane- and Bulk-MS in the cases of all $n$ 
(see Fig.\ref{fig:M-surface}). 
We also confirmed numerically that the inequality 
(\ref{eq:hoop-criterion}) holds in the cases of $n=1$ and $n=2$, 
while it does not in the case of $n=3$.
Then we can say that 
{\it
the Bulk-MS can form even for so highly elongated matter
distribution that the inequality $C>4\pi GM$ 
is satisfied.
} 

In Fig.1 the Bulk-MS does not agree 
with the Brane-MS at the brane in our example. 
The same result is also obtained in Ref.\cite{Ref:SS}.
To consider this behavior of the Bulk- and Brane-MS, it is
instructive to compare with the analytic solution (\ref{eq:V-solution}). 
In this solution, also, the Bulk-MS
($r=G_{5}\sigma/3$) does not agree with the Brane-MS
($R=2G_{5}\sigma/3$) at the brane ($\xi=0$). 
On the other hand, in the case of the static black string
solution Sch$\times${\bf R}\cite{Ref:GL}, the intersection of 
Bulk-MS with the brane agrees with the Brane-MS. 
This difference suggest the existence of gravitational
waves on the initial data considered here.

The existence of the cylindrically symmetric marginal surface 
shows that the inequality $C\lesssim 4\pi GM$ is not a
necessary condition for the formation of black holes with
horizons in the RS scenario.
This inequality will be just the sufficient condition for the
black hole formations in the RS scenario.
This suggests that massive spindle singularities in the RS brane 
is enclosed by event horizons if the existence of 
future null infinity and the global hyperbolicity of its 
causal past are guaranteed, while those in 4-dimensional 
general relativity do not. 
We say that the hoop conjecture, which seems to be natural in
the original 4-dimensional Einstein gravity, does no hold in the
RS brane world. 
We must note that this conclusion is not the extrapolation from
the linear analyses but the result based on the general
relativistic treatment without any weak field assumption.
Using this treatment, we have obtained the qualitatively same
result as the speculation based on the linear analysis.

From physical viewpoint, the existence of naked singularities
must be interpreted as the appearance of some new fundamental
physics.
In this sense, for the gravitational collapse of matter 
elongated sufficiently, 
the hoop conjecture in the original 4-dimensional
Einstein gravity can be used to judge the
appearance of the new fundamental physics. 
As discussed here, the formation of the highly elongated
marginal surfaces which violates the 4-dimensional hoop
conjecture means the ``signal'' of the RS brane world. 
Probably, this ``signal'' may be not so significant in the
astrophysical phenomena like collapse of the cylindrically
distributed cosmic dust, because the stringlike dust
concentration with the thickness less than $l$ is necessary to
form highly elongated marginal surfaces.
However we may expect that such phenomena happened on the cosmic
strings formation in some era of very early universe (Davis
treated early cosmic strings in the RS scenario within the
linear analysis~\cite{Ref:Davis}).  
If so, the scenario of universe evolution may be modified by 
considering such phenomena, 
and in the future observational data of the Universe, 
we will find the ``signal'' which shows the possibility of 
the RS brane world.

In this paper we presented only one important result on the hoop 
conjecture in the RS brane scenario. 
We may expect other appropriate criteria for the black hole 
formation in the RS scenario.
For further understanding and application of this result, more
detailed investigations should be done.
We should clarify the dependence on ${G_{5}\sigma/l}$ of the
formation of marginal surface to apply the result for cosmology, 
for example.  
We also note that our result does not mean necessarily that no
naked singularity forms in the RS scenario in its original
sense. 
The pancake-type gravitational collapse might lead to serious
naked singularities in the RS model, although it is not so
serious in the framework of 4-dimensional general relativity.
The ``signal'' of the appearance of the new fundamental physics
would be the pancake-type singularities on the brane 
rather than the stringlike matter distributions discussed here.
These issues are future problems and will be discussed elsewhere.

\noindent
{\bf Acknowledgement}

We would like to thank M.~Yamaguchi and N.~Kitazawa for
helpful discussions. We are also grateful to D.~Ida, A.~Ishibashi, 
H.~Ishihara, T.~Tanaka, G.~Uchida and H.~Kodama for their useful
comments and discussions. 
We acknowledge M.~Arima and H.~Kobayashi for their help in constructing the numerical code. 
One of the authors(T. M.) also wish to thank 
the financial support by Nihon University Grant for 2001.

This work was partially supported by the Grant-in-Aid 
for Creative Basic Research (No.~09NP0801) from the Japanese 
Ministry of Education, Science, Sports and Culture.


%
\end{multicols}

\begin{thebibliography}{00}
\bibitem{Ref:yamaguchi-kitazawa}
  M.~Yamaguchi, N.~Kitazawa, private communications.
  $Z_{2}$-symmetry is often imposed in higher dimensional models
  in particle physics to avoid a problem of the chirality of the
  matter fields on the brane.
\end{thebibliography}
\end{document}